\documentclass[reprint,english,aps,prl,twocolumn, amsmath, amssymb, superscriptaddress]{revtex4}
\usepackage[T1]{fontenc}
\usepackage{subfigure}
\usepackage[latin1]{inputenc}
\usepackage{graphicx}
\usepackage{epsfig}
\usepackage{prettyref}
\usepackage{babel}
\usepackage{caption}
\makeatother

\begin{document}

\title{Observation of three-dimensional Dirac semimetal state in  topological insulator Bi$_2$Se$_3$}

\author{Devendra Kumar}
\email{deveniit@gmail.com}
\affiliation{UGC-DAE Consortium for Scientific Research, University Campus, Khandwa Road, Indore-452001, India}

\author{Archana Lakhani}
\affiliation{UGC-DAE Consortium for Scientific Research, University Campus, Khandwa Road, Indore-452001, India}

\begin{abstract}
The three dimensional (3D) topological insulators  are predicted to exhibit a 3D Dirac semimetal state in critical regime of topological to trivial phase transition. Here  we demonstrate the first experimental evidence  of 3D Dirac semimetal state in topological insulator Bi$_2$Se$_3$ with bulk carrier concentration of  $\sim$ 10$^{19}$ cm$^{-3}$, using magneto-transport measurements.  At low temperatures, the resistivity of our Bi$_2$Se$_3$ crystal exhibits clear Shubnikov-de Haas (SdH) oscillations above 6~T.  The analysis of these oscillations through Lifshitz-Onsanger and Lifshitz-Kosevich theory reveals a non-trivial $\pi$ Berry phase coming from 3D bands, which is a decisive signature of 3D Dirac semimetal state. The large value of Dingle temperature and natural selenium vacancies in our crystal suggest that the observed 3D Dirac semimetal state is an outcome of enhanced strain field and weaker effective spin-orbit coupling.
\end{abstract}

\maketitle

The 3D topological insulators are an exotic class of quantum materials having insulating bulk and novel spin-momentum locked dissipationless conducting surface. The surface states of these materials consist of odd numbers of massless Dirac fermions with linear dispersion protected by time reversal symmetry~\cite{Hasan, Qi}. In 3D topological insulators, the chalcogenide Bi$_2$Se$_3$ has drawn ample attention because of relatively simple and robust surface states consisting of single Dirac cone at $\Gamma$ point of band structure and a large topological band gap (0.3 eV) between the bulk states~\cite{Zhang}. The existence of topological surface states in Bi$_2$Se$_3$ has been confirmed through angle resolved photoemission spectroscopy (ARPES)~\cite{Xia}, scanning tunneling microscopy~\cite{Cheng}, and transport measurements~\cite{Analytis1}. The as grown crystals of Bi$_2$Se$_3$ are typically n-type because of electron doping due to natural selenium vacancies~\cite{Cheng, Hor}. Therefore, the transport properties of Bi$_2$Se$_3$ are generally dominated by bulk conduction, for example, the temperature dependence of resistivity is metal like~\cite{Hor, Analytis, Butch, Eto} and  Shubnikov-de Haas (SdH) oscillations in resistivity have the characteristic of bulk Fermi surface~\cite{Analytis, Butch, Eto}. The bulk carrier concentration (n) and Fermi energy in Bi$_2$Se$_3$ can be tuned by either controlling the natural Se vacancies~\cite{Analytis, Butch}, gate-voltage~\cite{Checkelsky}, or chemical doping~\cite{Analytis1, Hor, Analytis, Ren}. For n$\sim$10$^{17}$cm$^{-3}$, SdH oscillations are from two dimensional (2D) states indicating the surface transport~\cite{Analytis1}, while for n>10$^{17}$cm$^{-3}$, SdH oscillations come from 3D states indicating the dominance of bulk conduction~\cite{Eto, Analytis, Butch}. For high n i.e. n$\sim$10$^{19}$cm$^{-3}$, bulk dominated transport with 2D like SdH and quantum Hall effect is reported~\cite{Cao}.

The Dirac fermions at the surface of 3D-topological insulators are two dimensional i.e. graphene like, and very recently, their three dimensional analog have been discovered in a new kind of Dirac material known as 3D  Dirac semimetal. In 3D Dirac semimetals, the bulk conduction and valance band touch at the discrete Dirac points and disperse linearly along all momentum directions~\cite{Young, Yang, He}.  The 3D Dirac semimetals are predicted to exist at the critical point of topological phase transition from topological to trivial -insulator  in a space inversion symmetry~\cite{Murakami}.  Such a topological phase transition can be realized by closing the bulk band gap at the critical point by tuning of effective spin-orbit coupling with chemical doping~\cite{Xu, Sato, Brahlek, Vanderbilt} or varying the lattice parameters through strain/pressure~\cite{Liu, Liu1, Young1}. Experimentally the 3D Dirac semimetal state has been observed by ARPES measurements on BiTl(S$_{0.5}$Se$_{0.5}$)$_2$~\cite{Xu}, (Bi$_{0.94}$In$_{0.06}$)$_2$Se$_3$~\cite{Brahlek}, and in TlBiSSe through transport measurements~\cite{Novak}.

In this letter we report the quantum transport evidence of 3D Dirac semimetal state in as prepared  Bi$_2$Se$_3$ single crystal. We show that the SdH oscillations in Bi$_2$Se$_3$ crystal having carrier concentration of $\sim$1.13$\times$10$^{19}$ cm$^{-3}$, originate from 3D bands and has a non-trivial $\pi$ Berry phase. The presence of $\pi$ Berry phase from 3D bands suggest that the electrons encircle the 3D Dirac point in the momentum space. The large Dingle temperature in our crystal indicates the presence of high strain field. This transformation from topological insulator to Dirac semimetal state is a result of enhanced strain field and weaker effective spin-orbit coupling.

\begin{figure}[]
\begin{centering}
\includegraphics[width=0.95\columnwidth]{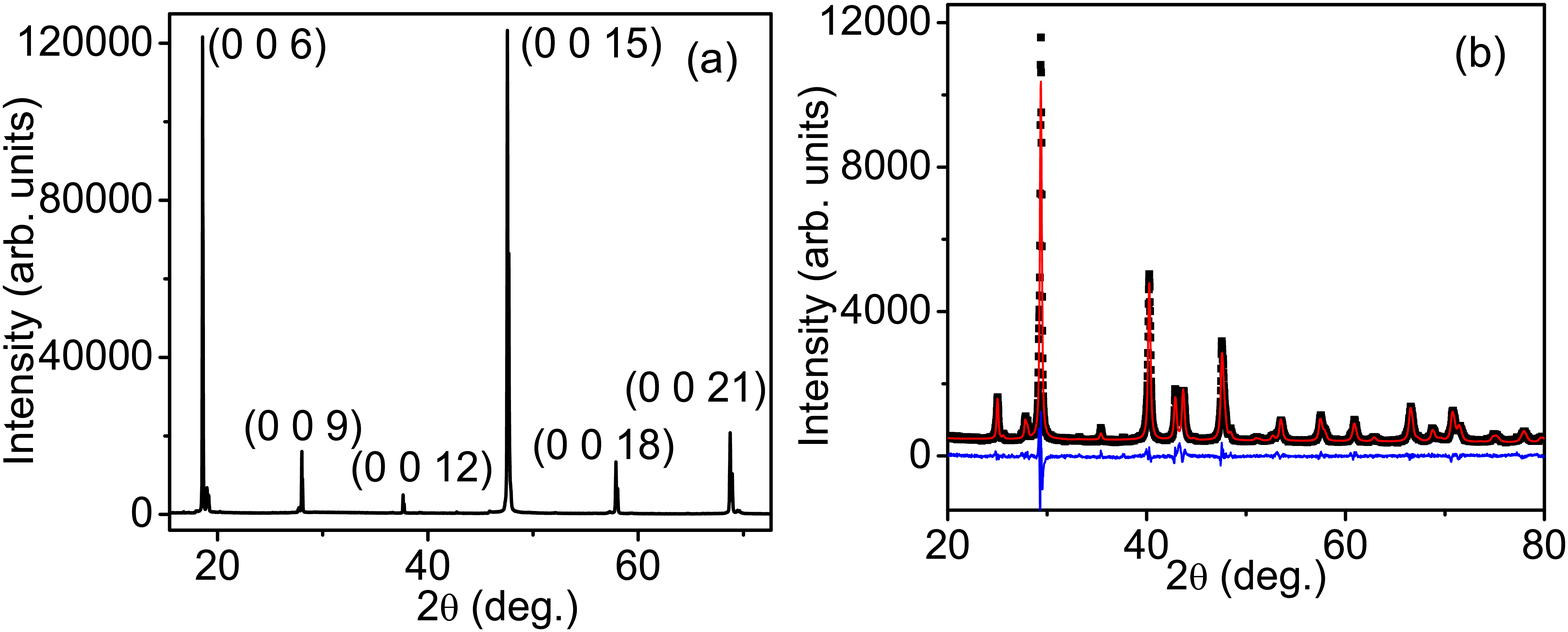}
\par\end{centering}
\caption{X-ray diffraction pattern of (a) freshly cleaved Bi$_2$Se$_3$ single crystal and (b) Bi$_2$Se$_3$ powder obtained after crushing the crystal. The symbols correspond to experimental data, red line represents the Le Bail fitting ($R3\bar{m}$ space group) and blue line is the difference between fitted and experimental data.}
\label{fig: XRD}
\end{figure}

Bi$_2$Se$_3$ single crystal is grown by modified Bridgeman method as in Reference~\cite{Cao}. The as grown crystal is easily cleaved which yields a flat shiny surface. The crystal is characterized by X-ray diffraction (XRD) performed on a Bruker D8 Advance X-ray diffractometer using Cu K$\alpha$ radiation, the data is shown in Figure~\ref{fig: XRD}~(a). The XRD data exhibit peaks only corresponding to (003) family of planes, suggesting the cleaved surface is perpendicular to C$_3$ axis of rhombohedral structure. The powder X-ray diffraction of the crushed crystal is displayed in  Figure~\ref{fig: XRD}~(b). The Le Bail Fitting of the powder XRD data gives $a$=4.144~\AA ~and $c$=28.660~\AA.  The energy dispersive X-ray analysis (EDX) yields atomic ratio of Bi and Se as 2 and 2.84 respectively.

\begin{figure}[]
\begin{centering}
\includegraphics[width=0.95\columnwidth]{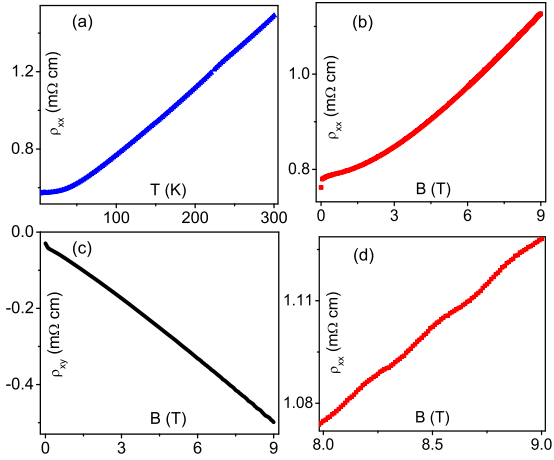}
\par\end{centering}
\caption{(a) Temperature dependence of longitudinal resistivity ($\rho_{xx}$) in zero magnetic field. (b) $\rho_{xx}$ and (c) Hall resistance ($\rho_{xy}$) as a function of magnetic field ($B\|$C$_3$ axis) at 1.8~K. (d) Expanded view of $\rho_{xx}$ versus $B$ plot exhibiting clear Shubinikov-de Haas oscillations.}
\label{fig: RT_RH}
\end{figure}

\begin{figure}[t]
\begin{centering}
\includegraphics[width=0.8\columnwidth]{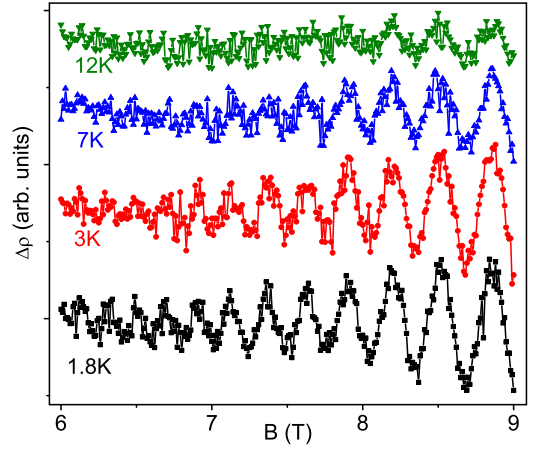}
\par\end{centering}
\caption{Oscillatory component of longitudinal resistivity ($\Delta \rho_{xx}$) as a function of magnetic filed at different temperatures. $\Delta \rho_{xx}$ is obtained after subtracting a smooth background from $\rho_{xx}$.}
\label{fig: TSdH}
\end{figure}

A thin rectangular bar shaped Bi$_2$Se$_3$ crystal was taken and contacts for resistivity and Hall measurements were made on freshly cleaved surface.
Resistivity measurements are performed using standard four probe ac technique while Hall measurements are carried through five probe ac technique on a 9T PPMS AC Transport system (Quantum Design). The angle dependent resistivity measurements are performed on the high resolution rotator option of PPMS.

Figure~\ref{fig: RT_RH}~(a) shows the temperature dependence of resistivity $\rho_{xx}$. The temperature variation of  resistivity displays metallic character with residual resistivity ratio (RRR) of about 2.7 which lies in the range of $\sim$1.6-3.0 as observed in earlier reports~\cite{Cao, Eto}. Figure~\ref{fig: RT_RH}~(b) and (c) display the field dependence (field parallel to C$_3$ axis) of longitudinal resistivity ($\rho_{xx}$) and Hall resistivity ($\rho_{xy}$) at 1.8~K respectively. The longitudinal resistivity ($\rho_{xx}$) increases on increasing the magnetic field, and above 6T, exhibits periodic oscillations with field. An expanded view of these oscillations is shown in figure~\ref{fig: RT_RH}~(d). The resistivity oscillations in metals occur due to formation of Landau levels at high magnetic fields and are known as Shubnikov-de Haas (SdH) oscillations. The result of field variation of Hall resistivity shows that our sample is n-type and a linear fitting of the data above 4~T gives  bulk carrier density  $n_{3D}^{Hall}$=1.13$\times$10$^{19}$ cm$^{-3}$. The Hall mobility ($\mu_H$) estimated from ratio of Hall coefficient and longitudinal resistivity is $\sim$986 cm$^2$Vs$^{-1}$. The n-type doping in  Bi$_2$Se$_3$ occurs due to natural Se vacancies.

\begin{figure}[!t]
\begin{centering}
\includegraphics[width=0.8\columnwidth]{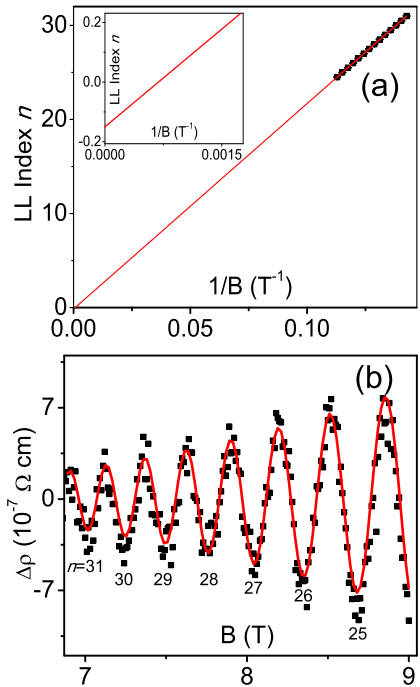}
\par\end{centering}
\caption{(a) Landau index $n$ versus 1/B plot for $\Delta \rho$  at 1.8~K. The straight line is the least square fitting of Lifshitz-Onsanger equation giving $\gamma$=0.15(5). Inset shows the magnified view around the intercept. (b) Direct fitting of Lifshitz-Kosevich equation to $\Delta \rho$  at 1.8~K. }
\label{fig: BerryPhase}
\end{figure}

Figure~\ref{fig: TSdH} displays the SdH oscillations in $\Delta \rho$ ($\Delta \rho$ is obtained after subtracting a polynomial background fit from $\rho_{xx}$) above 6~T in the temperature range of 1.8 to 12~K. The amplitude of SdH oscillations decreases on increasing the temperature and the oscillations are not resolvable above 12~K. The oscillations have a single frequency $B_F$=219(1)~T which is large in comparison to the previously reported values for Bi$_2$Se$_3$ single crystals~\cite{Eto, Analytis, Cao} and nano-flakes~\cite{Yan, Yan1, Kong}.  The Landau level index $n$ of the SdH oscillations is related to  extremal cross section area of Fermi surface ($\pi k^2_F$) perpendicular to the applied field through Lifshitz-Onsanger equation
\begin{equation}
2\pi(n+\gamma)=\pi k^2_F\frac{\hbar}{eB},\label{eq:LO}
\end{equation}
where $\hbar$ is the reduced Plank's constant, $e$ is the electronic charge, and $B$ is the applied magnetic field. The coefficient $\gamma$, which is equal but opposite in sign to the intercept on Y-axis in $n$ versus $1/B$ plot, is related to the phase of SdH oscillations. In absence of intraband and interband magnetic breakdown, for conventional metals $\gamma$=1/2~\cite{Shoenberg}. $\gamma$ gets an additional non-trivial contribution from Berry phase when electron orbit in $k$ space surrounds a contact line of its band with some other band, and with Dirac like dispersion, these band contact lines form the Dirac points~\cite{Mikitik}. In general, the phase of SdH oscillation $\gamma=1/2-\Phi_B/2\pi-\delta$, where $\Phi_B$ is the Berry phase ($\Phi_B$=0 for conventional metals and $\pi$ for Dirac systems) and $2\pi\delta$ is the phase shift correction determined by the dimensionality of Fermi surface~\cite{Murakawa, Zhao}. $\delta$=0 for 2D Fermi surface and $\delta$=$\pm1/8$ for 3D Fermi surface with positive sign for holes and  negative for electrons. Figure~\ref{fig: BerryPhase}~(a) exhibits the Landau level fan diagram, i.e. Landau level index $n$ versus 1/$B$ plot, for SdH oscillations at 1.8~K. The  minima of $\Delta \rho$ versus $B$ plot are assigned integer values of $n$  while half integer values $n$+1/2 are assigned to maxima~\cite{Qu, Barua}. The data in figure~\ref{fig: BerryPhase}~(a) fits well with  the equation~\ref{eq:LO} giving $\gamma$=0.15(5) and $k_F$=0.0815(1)~\AA$^{-1}$. $\gamma$ should be 1/2 for conventional metals, 0 for 2D and $\sim$1/8 (0.125) for 3D Dirac electrons with $\pi$ Berry phase. Therefore, $\gamma$=0.15(5) gives strong evidence for the presence of non-trivial $\pi$ Berry phase arising from 3D Dirac electrons. The $\gamma$ value is further reassessed by fitting the $\Delta \rho$ oscillations directly with Lifshitz-Kosevich equation~\cite{Tang, Barua}
\begin{equation}
\Delta \rho(B) = \Delta\rho_0(T) \text{exp}(-\pi/\mu B)\text{cos}2\pi(B_F/B+\gamma).\label{eq:LKeq}
\end{equation}
The fitting of equation~\ref{eq:LKeq} to SdH oscillations at 1.8~K is shown in Figure~\ref{fig: BerryPhase} (b).  We obtain fitting parameters $\gamma$=0.2(1), $B_F$=220(1), and $\mu$=8(1)$\times$10$^3$ cm$^2$/Vs. The $\gamma$ value from direct fitting reconfirms the result of Landau level fan diagram. SdH oscillations give carrier concentration of 5.3$\times$10$^{12}$ cm$^{-2}$ for 2D Fermi surface and 1.8$\times$10$^{19}$ cm$^{-3}$ for 3D Fermi surface. The mobility and carrier concentration (for 3D Fermi surface) values obtained from SdH oscillations are close to their corresponding values from Hall measurements which further support the origin of SdH oscillations from 3D bands. The relatively large value of $B_F$ (and $k_F$) in our system indicates that the SdH oscillations are coming from a large Fermi surface.

\begin{figure}[!t]
\begin{centering}
\includegraphics[width=0.8\columnwidth]{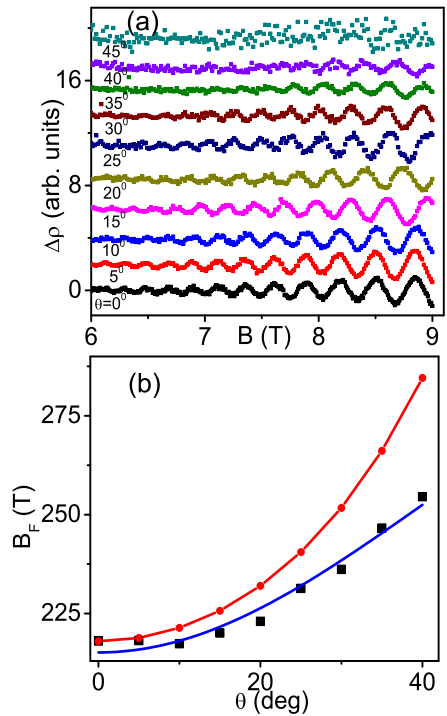}
\par\end{centering}
\caption{(a) SdH oscillations in $\Delta \rho$ at various tilt angles ($\theta$) at 2~K.  (b) The frequency of SdH oscillations B$_F$ as a function of tilt angle (Black squares). The red line with circles is the expected angle dependence of B$_F$ for 2D Fermi surface (B$_F\propto$ 1/cos$\theta$). The blue solid line is the least square fitting of prolate spheroid Fermi surface to B$_F$  versus $\theta$.}
\label{fig: AngleDep}
\end{figure}

The angle dependence of the SdH oscillations is used to determine the shape and size of the Fermi surface.  Figure~\ref{fig: AngleDep}~(a) shows the SdH oscillations at different tilt angles ($\theta$) while figure~\ref{fig: AngleDep}~(b) exhibits the angle dependence of SdH oscillation frequency  at 2~K. Here $\theta$ is the angle between C$_3$ axis and magnetic field; $\theta$ is zero when magnetic field is along the C$_3$ axis and  90$^0$ when magnetic field is perpendicular to the C$_3$ axis.  For 2D Fermi surface, the SdH oscillation frequency $\propto$ 1/cos$\theta$. The data shown in figure~\ref{fig: AngleDep}~(b) strongly deviates from 1/cos$\theta$ dependence which confirms  that the SdH oscillations in our system are not from 2D surface states.
The angle dependence of SdH oscillation frequency fits reasonably well with prolate spheroid Fermi surface  $B_F$=($\hbar^2 k_y^2/2e$)$\sqrt{(k_z^2/k_y^2)\text{sin}^2\theta+ \text{cos}^2\theta}$ with major axis $k_z$=0.112(2)~\AA$^{-1}$ and minor axis $k_x$=$k_y$=0.081(2)~\AA$^{-1}$.  These values of $k_x$ and $k_y$ are almost same as that reported for topological surface states in ARPES measurements ($k_y$=0.09~\AA$^{-1}$ and $k_x$=0.10~\AA$^{-1}$)~\cite{Xia}. The angle dependence of SdH oscillations  also confirms the origin of SdH oscillations in our system is from 3D bands.

\begin{figure}[!t]
\begin{centering}
\includegraphics[width=0.95\columnwidth]{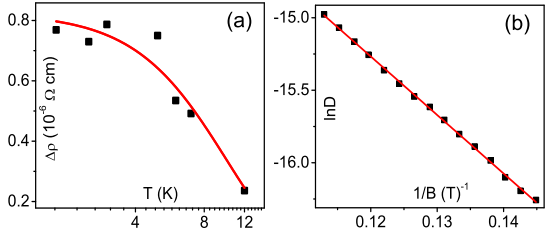}
\par\end{centering}
\caption{(a) The temperature dependence of SdH oscillation amplitude $\Delta \rho$ for $n$=25+1/2. The solid line is a fit of  equation~\ref{eq:LK} which gives $m^*$=0.15$m_e$. (b) Dingle plot of ln$D$ versus 1/$B$ at 1.8~K.}
\label{fig: Effectivemas}
\end{figure}

Figure~\ref{fig: Effectivemas}~(a) displays the temperature dependence of the amplitude of SdH oscillation for $n$=25+1/2 (B$\approx$8.85~T). The amplitude of SdH oscillation reduces on increasing the temperature and totally diminishes above 12~K. According to Lifshitz-Kosevich theory, the temperature damping of the amplitude of SdH oscillation can be given as~\cite{Tang, Barua}:
\begin{equation}
\Delta \rho_0(T)=\Delta \rho_0(0)\frac{2\pi^2k_BT/\hbar\omega_c}{\text{sinh}(2\pi^2k_BT/\hbar\omega_c)} ,\label{eq:LK}
\end{equation}
where $\Delta \rho_0(0)$ is the oscillation amplitude at T=0, $k_B$ is the Boltzman constant, and $\omega_c$ is the cyclotron frequency. $\omega_c$=$eB/m^*$ where $m^*$ is the electron effective mass. The fitting of equation~\ref{eq:LK} to the data of Figure~\ref{fig: Effectivemas}~(a) yields $m^*$=0.15$m_e$, where $m_e$(=9.1$\times$10$^{-31}$~kg) is the free electron mass. The value of electron effective mass is in agrement with the previous reports~\cite{Analytis, Eto, Cao}.  The energy  difference between two consecutive Landau levels $\Delta E$=$\hbar \omega_c$ $\sim$ 6.8 meV for 8.85~T. Using the value of electron effective mass ($m^*$) and Fermi wave vector ($k_F$), the Fermi velocity $v_F$=$\hbar k_F/m^*$ is estimated as 6.3$\times$10$^5$ m/s. The value of Fermi velocity is similar to that reported for surface states in ARPES measurement($\approx5\times10^5$m/s)~\cite{Xia}.  The Fermi energy $E_F$ determined from  relativistic formula $E_F$=$m^*v^2_F$ $\sim$ 336~meV above the Dirac point.

The Dingle temperature ($T_D$) which accounts for the broadening of Landau levels due to electron scattering is given as $T_D$= $\hbar/2\pi k_B\tau_s$, where $\tau_s$ is the quantum life time of carrier due to scattering. Figure~\ref{fig: Effectivemas}~(b) displays the semilog plot of $D$=$\Delta \rho B$sinh($2\pi^2k_BT/\hbar\omega_c$) versus 1/$B$ at 1.8~K and the Dingle temperature is estimated from the slope of linear fit which is $\sim$40~K. From
Dingle temperature we get the carrier life time $\tau_s$$\approx$3$\times$10$^{-14}$~s. The Dingle temperature of our sample is large in comparison to the previously reported values for Bi$_2$Se$_3$ e.g. T$_D$=3.5~K for n$\approx$2.3$\times$10$^{17}$ cm$^{-3}$~\cite{Analytis1}, T$_D$=9.5~K for n$\approx$5$\times$10$^{18}$ cm$^{-3}$\cite{Eto}, T$_D$=4.0~K for n$\approx$2.3$\times$10$^{19}$ cm$^{-3}$~\cite{Analytis1}, and T$_D$=25~K for n$\approx$5$\times$10$^{19}$ cm$^{-3}$\cite{Cao}. The large value of T$_D$ (for similar carrier concentration) in our sample is an indication of the presence of relatively high strain fields~\cite{Shoenberg}. The strain field plays a vital role in determining the topological properties of Bi$_2$Se$_3$, and it has been observed that the strain field may drive the Bi$_2$Se$_3$ system through topological to trivial -insulator phase transition~\cite{Liu}.

The observation of $\pi$ Berry phase in 3D bands confirms the presence of Dirac point with 3D electron dispersion. This suggests that our Bi$_2$Se$_3$ system is in a 3D Dirac semimetal state. The 3D Dirac semimetal state in  Bi$_2$Se$_3$  can be realized on closure of the bulk topological band gap in the critical regime of topological phase transition. It has been shown that the bulk topological band gap in  Bi$_2$Se$_3$ can be reduced on lowering the effective spin-orbit coupling \cite{Xu, Brahlek} as well as on enhancing the strain field in the crystal \cite{Young, Liu}. In our system, atomic ratio of Bi:Se is 2:2.84 due to natural Se vacancies, suggesting the possibility of a slightly lower effective spin-orbit coupling than that of perfect stoichiometric crystal. Furthermore, the observation of large Dingle temperature in comparison to pervious reports asserts the presence of enhanced strain field in the system. These results imply that the 3D Dirac semimetal state in our system is an outcome of the closing of bulk topological band gap due to weaker spin-orbit coupling and enhanced strain field.

In summary, magneto-transport measurements on our Bi$_2$Se$_3$  single crystal give strong evidence of non-trivial $\pi$ Berry phase originating from 3D bands. The observation of non-trivial $\pi$ Berry phase from 3D bands suggests that our Bi$_2$Se$_3$ system is in a 3D Dirac semimetal state. The analysis of Dingle temperature and Bi:Se atomic ratio indicate that the transformation from topological insulator to Dirac semimetal state is because of enhanced strain field and reduced effective spin-orbit coupling. The revelation of Dirac semimetal state would provide a new insight in understanding the electronic properties of topological insulators and the tunability of topological phase without external doping, which preserves the high electron mobility, would offer interesting opportunities for future device applications.

\section{Acknowledgements}
We thank A. Tayal and M. Gupta for XRD measurements and D. M. Phase for EDX measurements.  V. Ganesan and A. Banerjee are acknowledged for support.




\end{document}